\documentstyle[prl,aps,epsf,multicol]{revtex}

\begin{document}

%
\newcommand{\fig}[2]{\epsfxsize=#1\epsfbox{#2}}
%
%
%
%
 \newcommand{\passage}{
         \end{multicols}\widetext\noindent\rule{8.8cm}{.1mm}\rule{.1mm}{.4cm}} 
 \newcommand{\retour}{
         \noindent\rule{9.1cm}{0mm}\rule{.1mm}{.4cm}\rule[.4cm]{8.8cm}{.1mm}%
         \begin{multicols}{2} }
 \newcommand{\unecol}{\end{multicols}}
 \newcommand{\deuxcol}{\begin{multicols}{2}}
%

\tolerance 2000

\title{Melting of two dimensional solids on disordered substrate}
\author{David Carpentier and Pierre Le Doussal}
\address{CNRS-Laboratoire de Physique Th{\'e}orique de l'Ecole Normale 
Sup{\'e}rieure, 24 Rue Lhomond, 75231 Paris}

\maketitle

\begin{abstract}
We study 2D solids with weak substrate disorder, using Coulomb gas
renormalisation. The melting transition is found to be replaced by a sharp
crossover between a high $T$ liquid with thermally
induced dislocations, and a low $T$ glassy regime with  
disorder induced dislocations at scales larger than 
$\xi_{d}$ which we compute ($\xi_{d}\gg R_{c}\sim R_{a}$, the Larkin and
translational correlation lengths).  
We discuss experimental consequences, 
reminiscent of melting, such as size effects in vortex
flow and AC response in superconducting films. 
\end{abstract}

\deuxcol


Although the phase diagram of the mixed state of high $T_c$
superconductors\cite{blatter94} 
has considerably evolved since their discovery,
many questions remain. In $d=3$ the low field region
has been proposed as the topologically
ordered Bragg glass phase\cite{giamarchi94}. The role of dislocations remains
unclear near melting or at higher fields. To study analytically the effect of
disorder on  defective solids, $d=2$ appears as a natural starting point, also
important for numerous experimental systems besides thin 
superconductor films\cite{huberman79,berghuis90,theunissen96,yazdani94}, such as Wigner 
crystals in heterojunctions\cite{exp-wigner} and on the surface of
Helium\cite{exp-electron}, magnetic bubbles arrays\cite{exp-bubbles}.
In the absence of disorder, the continuous melting of a 2d crystal occurs as
dislocation pairs unbind at $T_{m}^{0}$ and is well described by the KTNHY
theory \cite{kosterlitz73,nelson79}. But no equivalent theoretical description
of melting with substrate disorder exists, though it has been studied in many
experiments\cite{berghuis90,theunissen96,exp-wigner,higgins96}.

Progress was made for the simpler problem of crystals with structural (i.e
internal) disorder \cite{nelson83} where melting occurs at a temperature
$T_{m}(\sigma)$ shifted downwards by the disorder strength $\sigma$ (Fig.1).
As discussed in \onlinecite{giamarchi94},
\cite{nelson83} is equivalent to treating only
the long wavelength part $\sigma$ of the substrate disorder.
On the other hand an analytic RG study of the 2d Bragg glass including 
{\it pinning disorder $g$}, {\it i.e} short
wavelength disorder
(at the cost of excluding by hand dislocations) 
was performed recently \cite{carpentier97,carraro97}.
A complete treatment however should include dislocations. Indeed
the general theoretical belief based on qualitative arguments 
and simulations\cite{shi91,blatter-2D} is that no true
solid (neither a lattice nor a vortex glass \cite{blatter-2D})
exist in 2D in presence of disorder at $T>0$, and
thus there should be no true melting transition.
On the other hand signatures reminiscent of melting are
observed in various experiments\cite{yazdani94,theunissen96}.
This thus calls for further studies.

In this Letter we derive renormalisation group (RG) equations
describing a solid on a disordered substrate and allowing for dislocations.
They generalize the KTNHY equations to weak pinning disorder near
melting. They are obtained from the RG analysis, performed here for the first time,
of the generic {\it elastic vector electromagnetic coulomb gas} (VECG) with
disorder\cite{inprep}.  
Though the asymptotic RG flow is always towards strong coupling
(disorder or/and dislocation fugacity) several studies are still possible for
weak pinning disorder $g$ but arbitrary $\sigma$ (see
below) which strongly suggest that the pure melting 
transition is replaced by a sharp crossover between two very distinct regimes:
(i) at higher T a {\it weakly disordered liquid} where disorder has no effect
at 
large scale and dislocations are present on all length scales greater than
  $\xi_{+}^{0}(T)$ as in the pure case (ii) at lower $T$ a {\it quasi
Bragg glass regime}, where disorder disrupts the quasi-long range order
of the lattice beyond a length\cite{footnote5} $R_{a}\sim R_c$ and 
dislocations appear on scales larger than {\it a new length} $\xi_{d}(T)$. 
 From the RG, here we estimate $\xi_{d} (T)$ both for $T<T_{m}$ where
$\xi_{d}\gg R_{c}\gg a$, and in the crossover region.
 At scales $L\leq \xi_{d}$, the low $T$ defect free glassy regime is 
described by the theory of \cite{carpentier97}, despite its defective nature
in the thermodynamic limit. 
The sharp crossover predicted between the two regimes at weak disorder
is analyzed and argued to account for experimental observation of 
2D melting transition, and of the 2D peak effect.
\begin{figure}[thb] 	\label{fig1}
 	\centerline{\fig{6.5cm}{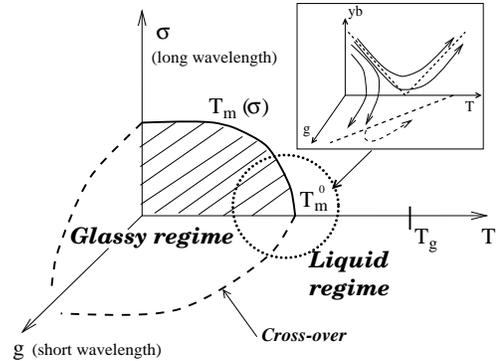}}
 	\caption{\narrowtext
        Schematic phase diagram: $\sigma$ and
       $g $ are respectively the long and short (pinning) wavelength
disorder. Inset: RG flow for weak $g $ and $T\sim T_{m}^{0}$. The
hatched region corresponds to the $g=0$ true solid which survives at $g=0$.}
\end{figure}
We emphasize that our RG equations
(5) are derived assuming that dislocations are thermalized. This 
assumption is reasonable  near $T_m$ and allows us to analyze
the crossover near melting.  A fully consistent RG analysis
including pinning of dislocations which become important at lower temperatures
\cite{giamarchi94}, is beyond the scope of this paper\cite{footnote:lowT}. However,
as a first step, we also studied the case of no pinning disorder
$g=0$ at $T\leq T_{m}^{0}/2$ by extending the analysis of 
\cite{nattermann95,scheidl96} on a simpler model.
A further study of melting is presented elsewhere\cite{inprep}.
 
In the absence of disorder and of dislocations a two dimensional crystal is
described by a smooth $2D$ displacement field ${\bf u}({\bf r})$ and
an elastic distortion energy:
\begin{equation}
        {\cal H}_{el}= \frac{1}{2} \int_{{\bf r}}
        2 c_{66}u_{ij}^{2}
        +(c_{11}-2c_{66}) u_{ii} u_{jj}
        \label{def-el}
 \end{equation}
where $c_{11}$ and $c_{66}$ are respectively the compression and the shear 
 elastic moduli, and
$u_{ij}=\frac{1}{2}(\partial_{i}u_{j}+\partial_{j}u_{i})$. 
For simplicity we study a triangular lattice of spacing $a$. Next, disorder
is modeled 
by a gaussian random potential $V({\bf r})$ with correlator
$\overline{V({\bf r}) V({\bf r}')} = h({\bf r}-{\bf r}')$ of range $r_f$, 
and couples to the density of vortices 
$\rho({\bf r})=\sum_{i}\delta({\bf r}-{\bf r}_{i})$
according to ${\cal H}_{p}= \int_{{\bf r}} \rho({\bf r}) V({\bf r})$. 
Decomposition of $\rho ({\bf r})$ and
of $V ({\bf r})$ in Fourier components gives several 
couplings between ${\bf u}({\bf r})$ and the disorder\cite{giamarchi94}. The
two main 
contributions\cite{footnote1} are a random stress field $\sigma_{ij}$ arising 
from the long wavelength part of the disorder, and a `random phase field'
$\phi_{\nu}$ which comes from the part of the (pinning) disorder with almost
the periodicity of the lattice:  
  \begin{eqnarray}\label{H-pinning}
       -\frac{{\cal H}_{p}}{T} & = & \int_{{\bf r}}   \sigma_{ij}u_{ij}
         + 2 \sqrt{\frac{g}{a^{2}}} \sum_{\nu=1,2,3} \cos \left(  {\bf
         K_{\nu}} .{\bf u}({\bf r}) + \phi_{\nu} ({\bf r}) \right)
  \end{eqnarray}
 with $\overline{ \langle e^{i(\phi_{\nu}({\bf r})-\phi_{\nu'}({\bf r}'))} 
        \rangle } = \delta_{\nu,\nu'} \delta^{2}({\bf r}-{\bf r}')$, the 
${\bf K}_{\nu}$ are the first reciprocal lattice vectors
(of length $K_{0}=4\pi/\sqrt{3} a$)
and the correlator 
$\overline{ \sigma_{ij}(r) \sigma_{kl}(r')}$ is parametrised
\cite{footnote:correl} by $\Delta_{11},\Delta_{66}$ whose bare values are 
$\Delta_{11}=\rho_0^2 h_{q=0}$, $\Delta_{66}=0$, while 
$g=a^{2} \rho_0^2 h_{q=K_0}/T^2$
where $\rho_0$ is the mean density. Note that
if $r_f \ge a$, $g$ can be {\it greatly reduced} (e.g. by a factor
$\exp(-c (r_f/a)^2)$) with respect to $\Delta_{11,66}$.
For out of plane disorder(such as in \cite{exp-electron}), varying
 the relative strength of the two types of disorder can be done by changing
the distance between the lattice and the disorder plane. 
 Besides this contribution, an underlying disorder potential induces 
local prefered orientation of the lattice\cite{footnote4} : this 
{\it new random field} (random torque) couples to the local bond
angle\cite{nelson79} 
$\theta=\frac{1}{2} (\partial_{x}u_{y}-\partial_{y}u_{x})$ : 
${\cal H}_{s}= \int_{{\bf r}} A ({\bf r}).\theta ({\bf r})$ with 
$\overline{A ({\bf r})A ({\bf r}')}=
 4 \Delta_{\gamma}\delta_{{\bf r},{\bf r}'}^{(2)}$. 
Even if $\Delta_{\gamma}$ is null in the bare model,
it will be renormalised to finite value (see (\ref{RG-equations})), and must
be taken into account from the beginning since it is a  
{\it new independent disorder strength}, whereas without
dislocations\cite{carpentier97} it can be absorbed in a
redefinition of 
$\Delta_{66}$ ($\Delta_{66}\rightarrow \Delta_{66}+\Delta_{\gamma}$). 

To describe plastic distortions of the lattice one 
splits ${\bf u}={\bf u}_{0}+{\bf u}_{d}$ into a smooth phonon part ${\bf
u}_{0}$ and a displacement field 
$u_{d,i}({\bf r}) = \frac{1}{2 \pi} \int_{\bf r'}
 {\cal G}_{ij}({\bf r}-{\bf r'}) b_j({\bf r'})$ due to edge dislocations of  
Burger vectors density ${\bf b}(r)$, where
 $$
 {\cal G}_{ij}({\bf r})=
 \delta_{ij} \Phi({\bf r}) + 
 \frac{c_{66}}{c_{11}} \epsilon_{ij} G \left(\frac{r}{a} \right) 
 +  \frac{c_{11}-c_{66}}{c_{11}} \epsilon_{jk} H_{ik}({\bf r}) 
 $$
 with $G({\bf r})+i \Phi({\bf r})=\ln(x+i y)$ where ${\bf r}=(x,y)$. 
The interaction $H_{ik}({\bf r})$ is defined by\cite{footnote7}  
$H_{ik}(r)=\frac{r_{i} r_{k}}{r^{2}}- \frac{1}{2} \delta_{ik}$.
Introducing $n$ replicas and averaging over disorder using the Villain
form for the cosine coupling in (\ref{H-pinning})
 we obtain\cite{carpentier97,inprep} in each site replicated
charges ${\bf m} = ({\bf m}_1,..{\bf m}_n)$, each ${\bf m}_a$
belonging to the reciprocal lattice, with the constraint\cite{carpentier97}
$\sum_{a} {\bf m}_a = {\bf 0}$ as well as the usual neutrality condition.
Integration over the replicated field ${\bf u}^a_{0}({\bf r})$
leads to an elastic VECG (instead of the purely electric studied in
\onlinecite{carpentier97}) which in its most general form is defined by 
$
\overline{Z^n}= 
\sum_{\{b,m \}} 
C^{-1} 
\prod_{\alpha=1}^{p}\int\frac{d^{2}{\bf r}_{\alpha}}{a^{2}}Y[{\bf
b}_{\alpha},{\bf m}_{\alpha}] 
e^{-S}
$
where the integrals are restricted by $|{\bf r}_{\alpha}-{\bf r}_{\beta}|\geq
a$, the 
$Y[{\bf b},{\bf m}]$ are the composite charges fugacities and $C(b,m)$ is a
combinatorial 
factor. The action is given by  
\begin{eqnarray}\label{action-CG}
 - &&S[b,m] = \frac{1}{2} \biggl\{ 
 {\bf b}^a*V^{ac}(\kappa_{1},\kappa_{2})*{\bf b}^c\\\nonumber &&+ 
 {\bf m}^a*V^{ac}(\kappa_{3},\kappa_{4})*{\bf m}^c\biggr\}
 +i ~{\bf m}^a*W^{ac}(\kappa_{5},\kappa_{6})*{\bf b}^c
 \end{eqnarray}
denoting 
${\bf b}*V*{\bf b}\equiv \sum_{ij}\int_{{\bf r},{\bf r}'}
{\bf b}_{i}({\bf r})V_{ij}({\bf r}-{\bf r}'){\bf b}_{j}({\bf r}')$. 
The interactions are $\pi V^{ac}_{ij}(\kappa_1,\kappa_2)
=\kappa^{ac}_{1} \delta_{ij}G -\kappa^{ac}_{2} 
  H_{ij}$ and $2 \pi W^{ac}_{ij}(\kappa_5,\kappa_6)=\delta_{ij} \delta^{ac}
\Phi +\kappa^{ac}_{5}\epsilon_{ij}G 
  +\kappa^{ac}_{6}\epsilon_{jk}H_{ik}$.
The renormalization of this {\it elastic} VECG 
goes beyond the previous analysis of the ECG \cite{nienhuis87}, due to the
presence of new  marginal
operators corresponding to the elastic interactions $V,W$.
Note that usual electric/magnetic 
self-duality for (\ref{action-CG}) now reads:
$
  \kappa_1  \leftrightarrow  \kappa_3  ~;~
  \kappa_2  \leftrightarrow  \kappa_4  ~;~
  \kappa_5   \leftrightarrow   -\kappa_5 ~;~
  \kappa_6  \leftrightarrow  \kappa_6
$
and exchanges direct and reciprocal lattices. The full RG equations 
and applications to various 2D models (depending on the definitions of the
$\kappa_{1\ldots 6}$) is presented elsewhere \cite{inprep}. 
Here we study the model (\ref{def-el},\ref{H-pinning})
for which the $\kappa_{1\ldots 6}$ are replica matrices of the form
$\kappa_{i}^{ac}=\overline{\kappa_{i}}\delta^{ac}-\Delta_{i}$
with the matrix definitions :
\begin{mathletters}
  \label{ref-VECG-constants}
  \begin{eqnarray}
  && \kappa_{1,2}=\frac{1}{T} 
	\left( (c_{11}-c_{66}) c_{66}c_{11}^{-1} \pm  
  \gamma c_{66} (\gamma +c_{66})^{-1}  \right) \\
  && \kappa_{3,4}= \frac{T}{4} \left( (c_{66}+\gamma)^{-1}\pm 
  c_{11}^{-1} \right) \\
  && \kappa_{5}=c_{66} c_{11}^{-1} - \gamma (\gamma +c_{66})^{-1}  
  ;\kappa_{6}-\kappa_{5}=1-2 c_{66} c_{11}^{-1} 
  \end{eqnarray}
\end{mathletters}
where the original parameters of the model (\ref{def-el},\ref{H-pinning})
have been embedded in three replica matrices
  $c_{11,66}^{ac}=c_{11,66}\delta^{ac}-\frac{1}{T} \Delta_{11,66}$
and $\gamma^{ac}=\gamma \delta^{ac}-\frac{1}{T}\Delta_{\gamma}$.
 Without coupling to a periodic lattice\cite{footnote4},
one sets $\gamma=0$.

This VECG can be studied by the RG as in \cite{nienhuis87}
by incrementing the hard-core cutoff $a\rightarrow ae^l$
in a three steps coarse-graining:
{\it reparametrisation} of the interaction, which gives the 
dimension of the operators : $\partial_l y
= ( 2-\frac{G^{2}}{2\pi } (\overline{\kappa}_1-\Delta_1 )) y,
\partial_l g
= ( 2-\frac{K_0^{2}}{2\pi } 2 \overline{\kappa}_3 )g$; 
{\it fusion} of electromagnetic charges separated from $a\leq d \leq ae^l$; 
{\it annihilation} of dipoles of diameter $a\leq d \leq ae^l$, 
 which defines scale dependent interactions $\kappa_{1\dots6}(l)$
\cite{kosterlitz73}. Here we restrict ourselves to charges of {\it minimal 
fugacity}: (i) charges ${\bf b}$ with single non zero
replica component $b_a={\bf G}$ (a lattice vector
of minimal length), $Y ({\bf b},0)=y $ (ii) charges ${\bf m}$ with two
(opposite) non zero replica component 
${\bf m}_a=-{\bf m}_c={\bf K}_\nu$, 
$Y (0,{\bf m})=g $\cite{footnote:lowT}. 
We show after rather tedious
calculations\cite{inprep} 
that (\ref{action-CG}) is renormalisable to lowest order
within the form (\ref{ref-VECG-constants}), needed for
consistency. We obtain
the general RG equations, which, in the case of
(\ref{def-el},\ref{H-pinning}), generalize the KTHNY equations: 
\begin{mathletters}
\label{RG-equations}
\begin{eqnarray}  \label{rg1}
& & \partial_l y  = \left[ 2-\frac{a^2}{8\pi}\left(
\frac{K}{T}-\frac{\delta+\sigma  K^{2}}{T^{2}} \right)\right]y 
 +2\pi \tilde{I}_{0} y ^{2} \\  \label{rg2}
& & \partial_l g  = \left[2-\frac{p^{2}
K_0^2}{4\pi}\frac{T}{c_{66}}\left(2-\frac{K}{4c_{66}}\right)\right]  
g -B_{m}(\alpha) g ^{2}\\
& & \partial_l K^{-1} = 
\frac{3\pi }{4 T}(2\tilde{I}_{0}-\tilde{I}_{1})y ^{2}  ~~,~~
 \partial_l c_{66}^{-1} = \frac{3\pi}{T} \tilde{I}_{0} y ^{2} \\
& & \partial_l \delta = 3\pi T^{2} p^{2} K_0^2 \left(I_{0} -I_{1} \right) 
g ^{2}  \\
& & \partial_l \sigma =  \frac{3\pi p^{2} K_0^2 T^{2}}{(4 c_{66} K)^{2}}
\left( 
(4c_{66}-K)^{2}(I_{0} +I_{1}) 
+ K^2 I_{0} \right)g ^{2}  
\end{eqnarray}
\end{mathletters}
where $I_{0,1}=I_{0,1}(\alpha)$ and 
$\tilde{I}_{0,1}=I_{0,1}(\tilde{\alpha})$ are modified Bessel function.
We have defined
$K=4 T \overline{\kappa}_{1}=4 T \overline{\kappa}_{2}=4 (c_{11}-c_{66})
c_{66}/c_{11}$, 
$\sigma K^{2}/2=T^{2} (\Delta_{1}+\Delta_{2})
=\Delta_{66} (1 - 2 c_{66}/c_{11}) 
+ \Delta_{11} (c_{66}/c_{11})^2$, 
$\delta=4 \Delta_{\gamma}$, and\cite{carpentier97} 
$B_{m}(\alpha) =2\pi(2I_{0} ( \alpha/2) - I_{0}(\alpha))$, 
$\tilde{\alpha}=\frac{a^2}{8\pi} (K/T-\sigma K^{2}/T^{2}+\delta/T^{2})$ and
$\alpha=p^{2}K_0^2 K T/(16\pi c_{66}^{2})$. We introduce an
additional parameter $p$ 
analogous to the p-fold symmetry breaking field in \cite{cardy82},
setting $K_0 \to p K_0$ in (\ref{H-pinning}). The physical
model corresponds to $p=1$.  We emphasize again that (\ref{RG-equations})
should be valid at high enough $T$ (near melting).

We start by the simpler case of {\it no
pinning disorder} $g=0$, where further extensions of (\ref{RG-equations}) at
low $T$ 
can also be given. The modifications induced by $g>0$ are discussed later for
$T>T_{m}/2$.
$g \simeq 0$ is experimentally relevant when disorder
($\Delta_{11,66,\gamma}$) 
varies very smoothly at the scale of the lattice.  At high enough temperature
$T> T_m^{(0)}/2$, eqs. (\ref{RG-equations})
show that the solid is stable at weak disorder
$\overline{\sigma} < \frac{1}{64 \pi}$ where 
$\overline{\sigma}a^2 =\sigma+\delta/K_{R}^2$    
and at low temperature 
$T<T_{m} (\sigma,\delta)=\frac{K_{R} a^2}{32 \pi} (1-\sqrt{1-64\pi
\overline{\sigma}}) $ 
(note that $\Delta_{11,66,\gamma}$ are unrenormalized). 
At $T=T_m(\sigma,\delta)<T_m^{(0)}$ it undergoes a 
true KT like melting transition (see Fig.1) to a high temperature phase where
dislocations proliferate. The correlation length at the transition is given
by $\xi_{+}^{0} \sim \exp({\rm const}(T-T_m)^{-\nu})$ where $\nu$, computed
in \cite{inprep} depends continuously
on $\sigma$ and $\delta$ and vanishes for 
$\overline{\sigma} = \frac{1}{64\pi}$.
When the random shear and torque are null  
$\Delta_\gamma=\Delta_{66}=0$, one recovers Nelson's results
\cite{nelson83}.  At lower temperatures ($T\leq T_m^0/2$),
 freezing of dislocations
leads to modified RG equations for 
${\cal T}\equiv T/(2 a^2 \overline{\sigma}K)<1$. 
We showed\cite{inprep} that it is possible to extend to the elastic model 
the approach of\cite{scheidl96} for the simpler scalar model,
by defining the appropriate  dislocation fugacities $\tilde{y}$ :
\begin{mathletters}\label{lowT}
\begin{eqnarray} 
&& \partial_l \tilde{y} = \left(2 - \frac{1}{32 \pi \overline{\sigma}}
\right) \tilde{y} ~~,~~ \partial_l \delta = 0 \\
&& \partial_l (TK^{-1}) = C (1-q) \tilde{y}^2 ~~,~~  \partial_l  \sigma = C q
\tilde{y}^2  
\end{eqnarray}
\end{mathletters} 
where $q \approx {\cal T}$ is the ratio of frozen dislocations\cite{f2}. 
The solid phase thus survives for $\overline{\sigma}_{R} < \frac{1}{64 \pi}$, as in
\cite{scheidl96}, and the physics is dominated by rare favorable regions 
for frozen dislocations \cite{inprep}, leading to the phase diagram of
Fig. 1.  Positional correlations decay algebraically with exponent
$\eta_G = \frac{G^{2}T}{4\pi}
(c_{11}^{-1}+c_{66}^{-1}+\frac{\Delta_{11}}{T c_{11}^{2}}+
\frac{\Delta_{66}+\Delta_{\gamma}}{T c_{66}^{2}})$, which leads at 
the transition $T= T_m^0/2$ to 
$\frac{1}{6}\leq \eta_{G} \leq \frac{1}{3}$ 
(for $\Delta_{66}=\Delta_{\gamma}=0$), and at $T=0$ to 
$\frac{1}{24}\leq \eta_{G} \leq \frac{1}{6}$ (depending on $c_{11}/c_{66}$).

We now turn to the effect of pinning disorder. Let us first
recall that in situations where dislocations
can be neglected (see below) one sets $y =0$ in 
(\ref{RG-equations}) and recovers
the Bragg glass phase\cite{carpentier97}
$g ^{*} B_{m}(\alpha)=2(1-T/T_{g})\equiv 2\tau $ which exists 
for $T\leq T_{g}=\frac{8 \pi c_{66}^{0}}{p^{2}} (2-K^{0}/ (4c_{66}^{0}))^{-1}$
(much larger than $T_m$). 
Due to the unbounded increase of $\Delta_{11,66} \sim l$, disorder
induced displacements grow as $u \sim \ln r$. The rest of the paper is devoted
to studying the situation 
where both weak pinning disorder and dislocations are included,
using eqs.(\ref{RG-equations}). The RG flow shows a sharp crossover between
two distincts regimes (see fig.2) : a 
 {\it high temperature regime} characterized by a correlation length
 $\xi_{+}^{0}(T)$ unaffected by pinning disorder 
({\it i.e} the same as given above), and a 
{\it low temperature glassy regime} where translational
order decays beyond $R_a \sim R_c$, though equilibrium
dislocations are separated (apart from small dipoles
of size $\sim a$) by the larger length
$\xi_{d}(T)$ (defined by $\xi_{d}=ae^{l^{*}},y(l^{*})\sim 1$). 
We also study the crossover by parametrizing
temperature and disorder 
using the correlation lengths $\xi_{\pm}^{0}(T)$ of melting in the absence of
pinning disorder and the Larkin
length in the absence of dislocations
$R^0_c \sim g^{-\frac{1}{2 \tau }}$ (see (35) in \cite{carpentier97}).
\begin{figure}[htb]\label{fig2}
	\centerline{\fig{8cm}{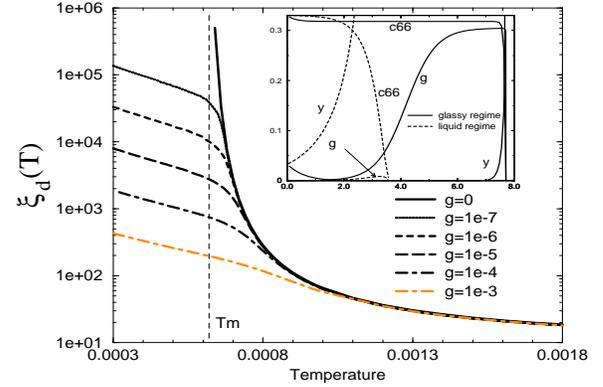}}
	\caption{\narrowtext 
	The length $\xi_{d}/a$ as a function of temperature, for
various 
pinning disorder strength $g $, and (rescaled) scale dependant parameters in
the two regimes (inset).}
\end{figure}
In the {\it low $T$ glassy regime} as $l$ increases $y (l)$ starts by
dropping quickly to a negligible value (Fig.2) whereas $g (l)$ increases and
reaches for $L = a e^l \gtrsim R_{c}\sim R_{c}^{0}$ 
a plateau value corresponding to the Bragg glass fixed point 
$g ^{*}$ defined above\cite{carpentier97,footnote:perturbative}.
Within this range of scale $K$ and $c_{66}$ remain constant since $y (l)\sim 0$.
However the steady growth of $\sigma$ and $\delta$ ($g (l)$ being finite) 
eventually makes the eigenvalue of $y$ positive.
Since $y (l)$ is then very small, it takes a large $l^{*}$ before
it reaches $\sim 1$ and  explodes to a non perturbative value.
Thus dislocations, which are strongly suppressed in the intermediate Bragg
glass regime, appear only on very large length scales: to estimate $\xi_{d}$, we can 
neglect the $y ^{2}$ terms in 
(\ref{RG-equations}) ($y  (l)\ll 1$ up to $\xi_{d}$). Integrating 
 the corresponding RG eqs., we obtain in the limit of weak disorder (large
$R_{c}^{0}$)
    \begin{equation}\label{xi-lowT}
    	\xi_{d}\sim R_{c}e^{b\sqrt{\ln R_{c}}} \text{ with } 
         b=2\tau \sqrt{\lambda (\sigma_{c}-\sigma_{0})}
    	\label{ref-def-xi}
    \end{equation}
\cite{f1} where $\lambda$ is a constant and 
$R_{c}\sim R_{c}^{0} (c_{66}(R_{c})/c_{66}(0))^{\frac{1}{\tau}}$ is the true Larkin length.
Note that $R_{c}\sim R_{c}^{0}$ in the low $T$ regime, except
very near the crossover region.

In the {\it high T liquid regime}, $y(l)$ increases quickly, 
causing $g (l)$ to decrease to $0$. Thus the correlation length 
is the same as without 
pinning disorder $\xi_{+}^{0}$. However this regime cannot be 
considerer as an hexatic phase: since the renormalised $\Delta_{\gamma}$ is
nonzero (computable from our RG), the  
corresponding random field destroys\cite{cardy82} 
long-range orientational order\cite{inprep,toner91} beyond $R_{H}\sim
(R_{c}^{0})^{2}/a$ (since  
$\Delta_{\gamma}\sim g^{2}$) leading to a {\it liquid} at large scales.
    
Finally, the crossover between the two regimes is dominated at weak enough
disorder by the $g=0$ fixed point (see fig.1).
 Studying the RG flow around the KTNHY separatrix\cite{nelson79}  
we find $\xi_{d}$ in the crossover region\cite{inprep} : 
\begin{equation}\label{xi-cross}
\xi_{d} \approx \frac{R_{c}^{0}}{(\ln R_{c}^{0})^{\frac{1}{4\tau }}} 
\left(1 \mp
\left(\frac{\ln R_{c}^{0}}{\ln \xi_{\pm}^{0}} \right)^{\frac{1}{\nu}} 
\right)^{\frac{1}{4\tau}}  
\end{equation}
where the $\pm$ sign depends on $T>T_{m}^{0}$ or $T<T_{m}^{0}$
\cite{footnote:crossover} and $\tau \approx 0.8$, $\nu$ as in \cite{nelson79}
near pure melting ($\sigma \approx 0$).
When increasing pinning disorder (or decreasing temperature), $\xi_{d}$ gradually 
goes from (\ref{xi-cross}) to the low $T$ behaviour (\ref{xi-lowT}). 

This sharp crossover will have consequences for finite size effects in 2D
systems such as thin  
superconducting films\cite{theunissen96,yazdani94,berghuis90}. 
In experiments of narrow channel vortex flow 
in Nb$_{3}$Ge$_{a}$ films\cite{theunissen96},
one probes the visco-elastic response of 
the lattice for $T<T_m$ 
on scales {\it much smaller} than $\xi_{d}$ (see fig.2) : 
the system responds as a solid with a finite shear modulus
$c_{66}$, and one observes a ``diverging'' correlation length (or
viscosity\cite{theunissen96}) when approaching $T_{m}(\sigma)$, 
as without pinning. In much larger system for $T\lesssim T_{m}$, the
response  
of the lattice should in fact be liquid-like with a large viscosity $\nu\sim 
\xi_{d}^{2}(T)$. 
In AC experiments\cite{yazdani94}, the 2D vortex lattice is probed 
on a length $l_{\omega}\sim \sqrt{D/\omega}$. By varying $\omega$,
a crossover at $l_{\omega_{c}}\equiv R_{\omega_{c}}$ was observed at 
low T in\cite{yazdani94}
between a low $\omega$ liquid-like behaviour and an activated glassy 
behaviour. The length $R_{\omega_{c}}$ strongly differs\cite{yazdani94}
from the  
corresponding $R_{c}$ extracted from critical current 
experiments. We argue that the observed 
$R_{\omega_{c}}$ of \cite{yazdani94} corresponds to the correlation length 
$\xi_{d}(T)$ defined here. Indeed the scaling of
$R_{\omega_{c}}$ with the sample 
thickness $d$ (i.e disorder strength) in \cite{yazdani94} is consistent
with (\ref{ref-def-xi}).  Finally, using (\ref{RG-equations}) we compute the
scale dependent $c_{66}(l)$ 
and obtain the softening of $c_{66}$ on scale
$R_{c}$ by dislocations. The self consistent
definition of $R_{c}$ given above then leads to
a quantitative description of 
the increase of critical current (peak effect) in 2D films\cite{higgins96}
near melting, further studied
in\cite{inprep}.  

To conclude we extended the KTNHY analysis in presence
of weak pinning disorder, predicted and analyzed a sharp crossover
near pure melting. To go beyond our study in the low $T$ region
would necessitate a controlled RG method\cite{inprep} to describe 
frozen topological defects.

\unecol

\begin{thebibliography}{99}
\bibitem{blatter94}G. Blatter et al., Rev. Mod. Phys, {\bf 66}, 1125 
(1994)
%
\bibitem{giamarchi94} T. Giamarchi and P. Le Doussal, 
Phys. Rev. B {\bf 52}, 1242 (1995)
%
\bibitem{huberman79}D. A. Huberman, and S. Doniach, Phys. Rev. Lett. {\bf 43},
950 (1979); D. S. Fisher, Phys. Rev. B {\bf 22}, 1190 (1980)
%
\bibitem{berghuis90} P. Berghuis et al, Phys. Rev. Lett. {\bf 65}, 2583 (1990)
%
\bibitem{theunissen96} M. H. Theunissen, E. van der Drift and P.H. Kes, 
Phys. Rev. Lett. {\bf 77}, 159 (1996), and T.H. Theunissen, PhD
%
\bibitem{yazdani94} A. Yazdani, Stanford PhD thesis (1994) and ref. therein
%
\bibitem{exp-wigner} E.Y. Andrei et al., Phys. Rev. Lett. {\bf 60}, 2765
(1988)
%
\bibitem{exp-electron} C.G. Grimes and G. Adams, Phys. Rev. Lett. {\bf 42},
795 (1979); 
F.I.B. Williams, Hel. Phys. Act. {\bf 65}, 297 (1992)
%
\bibitem{exp-bubbles} R. Seshadri and R.M. Westervelt, Phys. Rev. {\bf B 46},
5142 (1992); and {\it ibid} 5150 (1992).
%
\bibitem{kosterlitz73} J. Kosterlitz and D. Thouless, J. Phys C {\bf 
6}, 1181 (1973)
%
\bibitem{nelson79} D.R. Nelson and B.I. Halperin, Phys. Rev. {\bf B 19}, 
2457 (1979); A.P. Young, Phys. Rev. {\bf B 19}, 1855 (1979)
%
\bibitem{higgins96} M.J. Higgins and S. Bhattacharya, Physica C {\bf 
257}, 232 (1992) and references therein
%
\bibitem{nelson83}  D.R. Nelson, Phys. Rev. {\bf B 27}, 2902 (1983)
%
\bibitem{carpentier97}  D. Carpentier and P. Le Doussal, 
	Phys. Rev. {\bf B 55}, 12128 (1997)
%
\bibitem{carraro97} C. Carraro, D. R. Nelson, Phys. Rev. E {\bf 56},
	797 (1997) 
%
\bibitem{shi91} A. Shi and A. Berlinsky, Phys. Rev. Lett. {\bf 67}, 1926
	(1991) 
%
\bibitem{blatter-2D} see \cite{blatter94} section VIII.D.3
%
\bibitem{inprep}  D. Carpentier and P. Le Doussal, unpublished
%
\bibitem{footnote5}At such high temperature, $R_{c} \approx R_{a}$ (see
\cite{giamarchi94}) 

\bibitem{footnote:lowT}Higher charges
have higher naive dimensions, but composite charges {\it should} be included
 at low T\cite{inprep}.
%
\bibitem{nattermann95} T. Nattermann et al., J. Phys. I (France) {\bf 5},
 565 (1995); M. Cha and H.A. Fertig, Phys. Rev. Lett. {\bf 74}, 4867 (1995)
%
\bibitem{scheidl96} S. Scheidl Phys. Rev. B {\bf 55}, 457 (1997)
%
%
%
\bibitem{footnote1}Around
$T_{m}$ the only other relevant harmonic induces minor differences
%
\bibitem{footnote:correl} the correlator is given by
formula (7) in Ref. \cite{carpentier97} with the replacement
$\Delta_{11,66}$ of  \cite{carpentier97}  by $ \Delta_{11,66}/T$. 
%
\bibitem{footnote4} 
A periodic substrate (e.g. the atomic underlying lattice) induce a 
non zero $\gamma$ (see text). See the study in\cite{inprep}
%
\bibitem{footnote7} The 
$\delta_{ij}$ was absent in first
ref. of\cite{nelson79}: this only implies a different definition for  
the charge fugacities $y,g$. 
%
\bibitem{nienhuis87} B. Nienhuis, in {\it Phase transitions and 
critical phenomena}, C. Domb and J.L. Leibovitz Eds., {\bf 11} (1987)
%
\bibitem{cardy82} J. Cardy and S. Ostlund, Phys. Rev. B {\bf 25}, 6899 
(1982)
%
\bibitem{f2} 
After completion, we learned of a recent unpublished study
of the particular case $g=0$ (no pinning disorder (\ref{lowT})) by
P. Stahl, cond-mat/9710344 with similar results (though not identical
because of different RG schemes).
%
\bibitem{footnote:perturbative} The low $T$ regime can be studied
perturbatively beyond $R_c$ by
 choosing $p \approx p_c$ such that $T_{g}-T_{m}(\sigma)={\it 
 O}(p-p_c)\ll 1$, and extrapolate to $p=1$
 at the end.
%
\bibitem{f1}
This extends to triangular {\it elastic} lattices the estimate of 
\cite{giamarchi94} based on the simpler $N=1$ component model.
%
\bibitem{toner91} J. Toner, Phys. Rev. Lett. {\bf 67}, 1810 (1991)
%
\bibitem{footnote:crossover} (\ref{xi-cross}) is valid, {\it
e.g.} for $T > T_m^0$, 
for $R_c^0,\xi_{+}^0 \gg a$, and fixed $\alpha=\frac{\ln R_c^0}{\ln
\xi_{+}^0}$ ($\alpha < 1$). For $\alpha \gtrsim 1$, $\ln \xi_d \sim \ln
\xi_{+}^0$. 

\end{thebibliography}
\end{document}